\documentclass[conference, a4pape]{IEEEtran}

\usepackage{cite}
\ifCLASSINFOpdf
   \usepackage[pdftex]{graphicx}
\else
   \usepackage[dvips]{graphicx}
\fi

\usepackage[cmex10]{amsmath}
\usepackage{algorithmic}
\usepackage{algorithm}

\usepackage{array}
\usepackage{mdwmath}
\usepackage{mdwtab}
\usepackage{eqparbox}
\usepackage[tight,footnotesize]{subfigure}
\usepackage{fixltx2e}
\usepackage{stfloats}
\usepackage{url}

\begin{document}

\title{Fixing Data Anomalies with Prediction Based Algorithm in Wireless Sensor Networks\\}

\author{\IEEEauthorblockN{Abhishek Kr. Singh}
\IEEEauthorblockA{IIM Bangalore\\
Bangalore - 560076, India\\
Email: abhishek.singh@iimb.ernet.in}
\and
\IEEEauthorblockN{Bollibisai Giridhar}
\IEEEauthorblockA{Infosys Limited\\
Bangalore - 560100, India\\
Email: giridhar\_b@infosys.com}
\and
\IEEEauthorblockN{Partha Sarathi Mandal}
\IEEEauthorblockA{Indian Institute of Technology Guwahati\\
Guwahati - 781039, India\\
E-mail: psm@iitg.ernet.in}
}
\maketitle
\begin{abstract}
Data inconsistencies are present in the data collected over a large wireless sensor network (WSN),
usually deployed for any kind of monitoring applications. Before passing this data to some WSN
applications for decision making, it is necessary to ensure that the data received are clean and accurate.
In this paper, we have used a statistical tool to examine the past data to fit in a highly sophisticated
prediction model {\it i.e.}, ARIMA for a given sensor node and with this, the model corrects the data using 
forecast value if any data anomaly exists there. Another scheme is also proposed for detecting data anomaly at sink among the
aggregated data in the data are received from a particular sensor node. The effectiveness of our methods
are validated by data collected over a real WSN application consisting of Crossbow IRIS Motes \cite{Crossbow:2009}.
\end{abstract}

{\it Key words:} Anomalous data, Data forecasting, ARIMA model, Wireless sensor networks.

\ifCLASSOPTIONpeerreview
\begin{center} \bfseries EDICS Category: 3-BBND \end{center}
\fi
\IEEEpeerreviewmaketitle

\section{Introduction}
Wireless Sensor Networks (WSNs) are formed by large number of autonomous units called sensor nodes.
Each sensor node has the capability of sampling data, processing it and sending the data through radio
transmitters. In this aspect, each sensor node is independent of its sampling and sending mechanisms
and its values. This independent working nature of sensor nodes set up notion of independent data
transmission to the base station. The aggregated data at sink are independent. The base station
is a processing center also called sink node or simple sink.

WSNs are extensively used in natural environment monitoring and inventory management. Lots of specific
applications have been developed to monitor very delegate processes that include: nuclear reactor control,
habitat monitoring, object tracking, mines monitoring, fire detection, wild life monitoring, etc.
Depending on the application and user requirement, sensor nodes report the data to the sink either
in synchronous mode or in asynchronous mode. Usually sensor nodes sensed data in a
fixed time indexed manner and transmit the data to the sink periodically.

%We expect data from sensor nodes as discussed above to be reported readily in a fixed time indexed manner.
%It is very important for an application to make decisions on fly with the time interval of transmission reducing to fraction of seconds in many cases.

The WSNs based applications that we have mentioned above use aggregated data to perform a certain task
and give meaningful outputs to the network or to the user. The aggregated data from the WSN may be
affected by anomalies in the WSN. The anomaly detection is possible when the aggregated data at sink
do not follow a certain pattern \cite{Kaur:2010}. Anomalous data patterns can be caused due to
{\it case 1}: unreliability of wireless sensor networks or {\it case 2}: due to occurrence of unusual phenomena in the monitored region.
For {\it case 1}: the unreliability of wireless sensor networks incurs faulty sensors and the faults occur due to hardware malfunction,
sampling errors, transmission loss etc. Detecting data anomaly for both of the cases, {\it case 1 {\rm and} 2} are very important with respect to
any type of monitoring applications. One important objective of WSN application is to detect the occurrence
of unusual phenomena in the monitored region and to take necessary action for that. Another objective of WSN application
is to make appropriate decisions based on aggregated data at sink in spite of the unreliability of the wireless sensor networks.
Hence, it becomes crucial for us to correct the data before applying it to the applications.
Otherwise, the anomalous data produced due to unreliability of wireless network will have a
great impact on making appropriate decisions.

In this paper we make an attempt to exploit behavior of a single node over a considerable time to correct data
if there is any anomaly in the data. Specifically, we fit a statistical model to a single node as we know, all
nodes transmit data independent of each others and it is quiet clear that we may not know the spatial information
before hand. We validate our model with data gathered over a real WSN for considerable period of time based on
the IRIS platform \cite{Crossbow:2009}.

\subsection{Contributions}

In this paper we present an appropriate statistical modeling {\it i.e.}, $ARIMA(p,d,q)$ using the data of a real WSN application
consisting of Crossbow IRIS Motes. We propose an algorithm \ref{alg:1}: {\it To find suitable ARIMA model and Forecast},
which corrects the anomalous data at sink for each sensor node with ARIMA forecast values
at any point of time. The forecast values are also used in the algorithm \ref{alg:2}: {\it Anomaly Detection} for detecting
anomalous data of a sensor node with 95\% confidence interval.
The algorithms applied for each node are solely dependent upon the data stream transmitted by that particular sensor node.
As the algorithms use past data of individual node only, it is imperative that the algorithms do not depend
upon state of other nodes in the network. We also do not consider contextual and temporal relationship among the
nodes to predict the forecast value. While, if needed, contextual or temporal relationship can be used to further
smooth our results as suggested by \cite{Ali:2010}.

The advantages of the proposed works are following compare to the earlier works.
Our anomaly correction algorithm only needs data from the particular node we want to study.
The proposed algorithm can be used for the purpose of fault tolerance in the following way.
If few nodes fail to sense data due to transient fault at a particular instance of time, still we can produce data by
processing its old data. Our method is highly sophistic method, ARIMA, in statistics
time series models are known to represent many complex processes than any other models.
Once the preliminary condition of stationary is satisfied then we can use them to represent complex series.
Finally, all our data processing is to be done at the sink, which is suppose to have
sufficient power and enough computational capability for fitting the statistics models,
detecting and correcting the anomaly for the data of any sensor node.
%So, we can run the model to higher complexity also which in not possible to do in a sensor node.

\subsection{Related works}

Statistical modeling is used in literature for the purpose of data gathering with less number of
transmission, anomaly detection in the gathered data at sink etc.
When sampling of data is being done on regular time intervals, we get a time series. Time series is a
well researched topic in statistical mathematics field. An interesting observation has been done in
case of natural environmental data sampling by wireless sensor nodes, the time series is usually stationary in nature.
This property is used for the purpose of choosing a suitable statistical modeling, related papers are explained below.

A method is proposed by Liu {\it et al.} in \cite{Chong:2005} to reduce the transmission in the network.
The method uses ARIMA model to construct a prediction model for sampled data. Specifically, in this method the model
is being run on both sensor node and at the base station. If the difference between value sampled at sensor node and
the value forecasted by ARIMA model is smaller than a pre-defined tolerance, the value is not transmitted over network
to the base station. In this case base station is also running the same model, and hence use the forecasted value
as actual value. This method has shown suppression of data transmission upto 78\%. In this method time synchronization
is an big issue when the clocks running at node are different from the base station or time delays in reporting.
Jain and Chang in their paper \cite{Jain:2004} used recursive models to reduce transmission rate by setting the
transmission rate adaptively. In this paper \cite{Jain:2004}, the authors uses Kalman-Filter based estimation technique
to give new sample rate when required. Adaptively adjusting the sampling rate reduces transmission frequency which result
in energy saving. Anomalous behavior is common in WSNs. This fact puts forward a challenge to detect these anomalies
with high efficiency. This paper \cite{Sutharshan:2007}, Rajasegarar {\it et al.} have used Intel Berkeley Research
Laboratory (IBRL) data to suggest an anomaly detection method using statistical concept of Mahanbolis Distance.
The authors construct a hyperellipsoidal boundary using Mahanbolis distance, which provide the boundary for acceptance
or rejection of data as good or anomaly.
Ali {\it et al.} in the paper \cite{Ali:2010} directly deals with data cleaning for a set of aggregated data.
The technique takes into account the contextual association among sensor nodes. The method is named as Time of
Arrival Data Cleaning (TOAD). Depending upon belief range the authors proposed a scheme to select proper filter
and try to minimize the difference between the actual value in the environment and the data received at the sink.
Clustering can be used to club similar data points. In this paper \cite{Rajasegarar06distributedanomaly},
clustering is done based on the measure of dissimilarity among data. The concept of Euclidean distance between pair
is used to make clusters. The method uses Average Inter Cluster Distance (ICD) as the criteria for accepting or rejecting
anomaly cluster. Another similar works proposed by Chitradevi {\it et al.} in \cite{Chitradevi:2011} using the concepts of
clustering. In this paper a scalable cluster based anomaly detection algorithm is proposed by the authors,
where the algorithm locates anomalous clusters within sensor stream and enables the detection of both
local and global anomalies.

\section{Basic Idea}
We assume sensor nodes are deployed  over a region and formed a wireless sensor network.
Each sensor node is sampling data periodically. In our experiment sensor nodes are sensing temperature and light data,
and transmit data to the base station (sink) via multi-hop network. We use multi-hop wireless mesh network for our experiment.
The aggregated data store in a database at the sink for the further processing. Before processing further we should
identify all anomalous data.

In the context of this paper we observe two types of data anomaly. The first type is {\it irregular data}
which may generate due to occurrence or presence of unusual phenomena in the monitored region and
the second type is {\it erroneous data} which may generate due to faulty sensor and/or unreliability of wireless
communication during data transmission toward the sink. Considering the above two types of data, we define data {\it anomaly}
which is given below. When data do not follow stationary time series then we consider that there is an {\it anomaly}
of the data aggregated at the sink otherwise data are regular. Both types of data are important for the evaluation purpose
of a WSN application. In case of irregular data there might be a positive signal for occurrence of an unusual
phenomena otherwise the data are erroneous. In case of erroneous data we have to replace or correct the data with
an appropriate data which should follow stationary time series and remove the anomaly.

The goals of this paper are detecting the anomaly and fixing anomaly within the data to each node by ARIMA forecast \cite{Box:1994}.

\subsection{ARIMA Models}

The Auto-Regressive Integrated Moving Average (ARIMA) \cite{Box:1994} models are a class of models for
forecasting a time series. Unless a time series is stationary it is not possible to apply ARIMA models
for forecasting. The property of stationary time series is that over time, statistical properties like
mean, variance are constant. If the initial time series is non-stationary then by taking the differences
between successive values it is possible to make the series stationary. In practice first order difference
($X^1_t$) is used to make a time series ($T_t$) stationary, where $X^1_t=T_t-T_{t-1}$. The parametric
representation of $T_t$ is $ARIMA(p,d,q)$. The time series $T_t$ becomes stationary time series, $X^d_t$
after $d$ times differencing over itself for all $t$.  The $X^d_t$ is represented below.

$X^d_t = \phi_1 \times X^d_{t-1}+\cdots+ \phi_p \times X^d_{t-p}+ Y_{t}+ \theta_1\times Y_{t-1}+\cdots+ \theta_q\times Y_{t-q}$

Where $p$ is the number of auto-regressive (AR) terms and $q$ is the number of moving average (MA) terms.
With each AR term, there is an associated lagged dependent data sets, $X^d_{t-1}, \cdots, X^d_{t-p}$ and
for each MA term there is an associated random shocks,  $Y_{t}, \cdots, Y_{t-q}$ respectively. Examination
of the partial autocorrelation function (PACF) and the autocorrelation function (ACF) are required to get
an idea of what orders ($p$) to be considered for the AR component and what orders ($q$) to be considered for
the MA component respectively. For a given set of data funding appropriate values of $p$, $d$ and  $q$ is the main task to
fit an appropriate ARIMA model. After that the $ARIMA(p,d,q)$ can be used for the entire data set.

\section{Fixing anomalies within particular node}

Objective in this section is to fix anomalies data within particular node. The basic idea of time series modeling
is to make use of autocorrelation structure in the data sampled by the node in past.
Our method works in three phases: namely Data sampling, Applying statistical tests and Forecasting values
with suitable model.

{\it Data sampling}: In data sampling phase sensor nodes are working normally in supervision or sensed environmental
parameters like light and temperature periodically and transmit the sensed data to the sink.
Data are time indexed and stored at sink with equal time intervals. During this
phase sensor nodes sample sufficient data to build the statistical model $ARIMA(p,d,q)$
with appropriate values of $p$, $d$ and  $q$. As the amount of past data increases the model become more efficient
in forecast. In this phase, there is no need to maintain any extra database at sink excepting the aggregated data.

{\it Applying statistical test}: In this phase we start applying our statistical tests to check whether the data
sampled in previous phase qualify for time series analysis or not. Examine ACF for stationarity.
The ACF for a non-stationary series shows large autocorrelations that diminish only very slowly at large lags.
If sampled data are non-stationary then differences and further ACF examination is required until the transformed
time series become stationary. We collect the sampled data till the point of first anomaly reported in the data
stream and apply stationary time series tests. Simple way to analyze the autocorrelation structure is to plot
autocorrelation function (ACF) and partial autocorrelation function (PACF) with respect to different {\it lags}.

{\it Forecasting with suitable model}: The objective of finding suitable model is to determine the right order
of the AR component and the MA component respectively. The traditional criteria for ARIMA model selection are Akaike
Information Criterion  and Schwarz Criterion \cite{Box:1994}. As we are working at sink it is assumed that we have
enough computational power to run any criteria for model selection. For our purpose we have selected Akaike
Information Criterion (AIC). Applying AIC we can find out AR component order ($p$) and MA component order ($q$).
The $d$ in ARIMA stands for the number of times the data have been differenced to render to stationary.
And hence that will give us our suitable ARIMA model with $p$, $d$ and $q$ parameters.

Once the model selection is over, we can use the model for forecasting and detecting and correcting anomaly
with a forecast value. Usually five values are good to forecast at a single point, as more than five will
result in accumulated statistical error. It is also possible that for more than five consecutive anomalies
data for a sensor node, result some malicious activity or physical problems, so we restrict the forecasting
to maximum five steps. To continue after five steps we update the data and again start the whole process.

The following proposed Algorithm \ref{alg:1} can be used for finding appropriate value of the parameters $p$,
$d$ and $q$ which fit an $ARIMA(p,d,q)$ model, after that the model is used for forecasting future values and
fixing anomaly.
\begin{algorithm}
\caption{To find suitable ARIMA model and Forecast}
\label{alg:1}
\begin{algorithmic}[1]
\STATE Find autocorrelation structure existence, e.g. by PACF and ACF or lag plot
\STATE Fit a $AR(p)$, order $p$ calculated with AIC criterion
\STATE Fit the residual of step 2 into $MA(q)$ with order $q$
\STATE Make the residual analysis
\STATE Forecast future values and fix the anomaly
\end{algorithmic}
\end{algorithm}

The following proposed Algorithm \ref{alg:2} is used for detecting anomaly among the sensed data aggregated at
sink for a particular sensor node. The same algorithm is applicable for all sensor nodes.
\begin{algorithm}
\caption{Anomaly Detection}
\label{alg:2}
\begin{algorithmic}[1]
\STATE Forecast future values with ARIMA model generated in Algorithm \ref{alg:1}
\STATE Find the 95\% confidence interval as $\mu \pm 1.96 \sigma$ where $\mu$-forecast value, $\sigma$-standard error
\STATE Test the null hypotheses: Reported value lies in between the above interval.
\STATE Depending upon result of step 3, reject or accept reported value.
\STATE The anomaly is detected in case of rejection in step 4.
\end{algorithmic}
\end{algorithm}

\section{Experimental results}
\begin{figure}[h]
\centering
\includegraphics[width=3.3in]{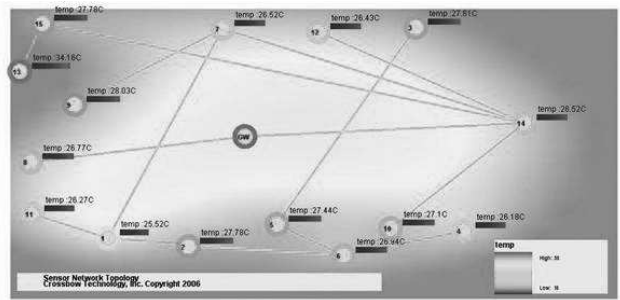}
\caption{Network Topology}
\label{fig_topology}
\end{figure}

Battery powered Crossbow IRIS mote platform is used in experimental setup.
A whole new setup of Xmesh
network was explored to setup a self
sustaining wireless network. Mote View software was used to monitor a network
deployed in real environment with 15 motes.
The motes were equipped with light sensors, temperature sensors, radio transceiver with ATMU 1281 microprocessor.
The motes are deployed in different locations considering the surrounding variation of temperature and light,
such as inside of a room
with AC and without AC, outside of the room, on a ladder to roof,
near AC room compressor etc. Topology of the setup is shown in the figure \ref{fig_topology}.

\begin{figure}[h]
\centering
\includegraphics[width=3.0in]{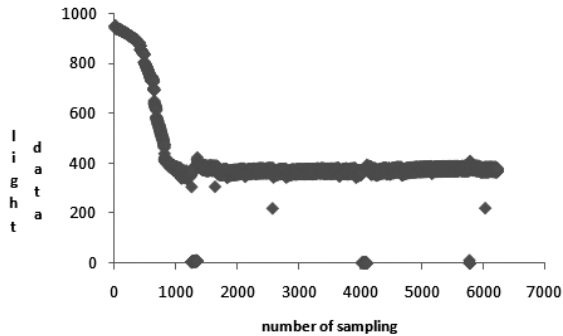}
\caption{Figure showing a scatter plot of the data received from mote 7}
\label{fig_scatter}
\end{figure}

\begin{figure}[h]
\centering
\includegraphics[width=3.0in]{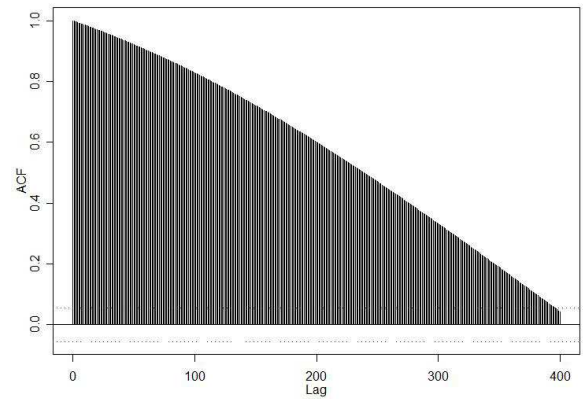}
\caption{Figure showing a plot of ACF versus Lags for the data of group 1}
\label{fig_acf_plot}
\end{figure}

The data were collected simultaneously for around 4 hours from all motes,
as we started our experiment at 4.27PM while sunlight was high and temperature was also high.
Then we stopped our experiment at 8.20 PM, meanwhile it was dark and temperature also dropped.
The motes were set on high power state to aggregate the data at base station.
The base station forwards the aggregated data to a Laptop.
The high power state makes network rearrangements at every 36 seconds and data
sampling at every 2 seconds. We collected data over network of
15 motes based on temporal and spatial variation.
We performed our mathematical analysis based on the received data from mote 7 (mote id).
The scatter diagram of the first 6200 sampled data of mote 7 is showing in the figure \ref{fig_scatter}.
\begin{figure}[h]
\centering
\includegraphics[width=3.0in]{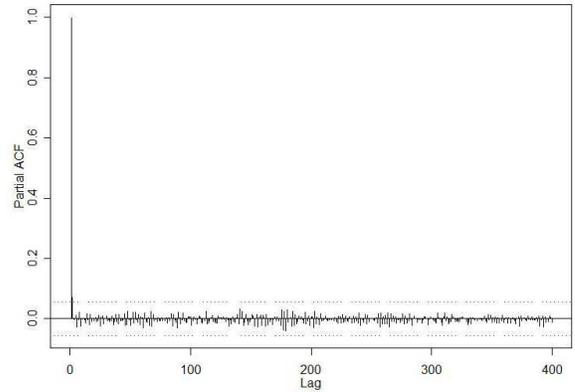}
\caption{Figure showing a plot of PACF versus Lags for the data of group 1}
\label{fig_pacf_plot}
\end{figure}
The scatter diagram indicates the presence of anomalous data, which are located isolated among the
sampled data of mote 7. The diagram also indicates a rapid drop of the data value at around 1000 sample,
which is due to the transition time of day and night. Here we have separated the data set into two groups
for analysis purpose. In group 1 we have taken sample from 1 to 1000 and in the group 2 we have taken
sample from 1500 to 4000. First we are analyzing the data of Group 1 as follows.
Sample data of group 1 is used to plot ACF verses lag and PACF verses lag.
The ACF verses lag and PACF verses lag plots give us an idea how well the data fit in ARIMA model.
The figure \ref{fig_pacf_plot} shows that PACF is decreasing fast and the figure \ref{fig_acf_plot}
shows that ACF is decreasing exponentially.
It implies that the sample data belonging to the group 1 are stationary.
Therefore, the statistical test ensures that the sample data of group 1 are going to fit an ARIMA model.
\begin{figure}[h]
\centering
\includegraphics[width=2.2in]{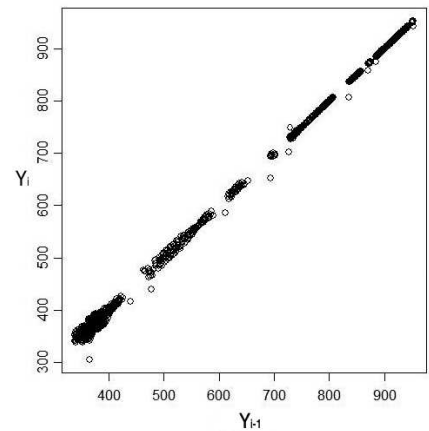}
\caption{Figure showing a Lag Plot for the data received from mote 7}
\label{fig_lag_plot}
\end{figure}

Another simple way to analyze the autocorrelation structure of the data set is Lag Plot.
A lag plot identify non-randomness of a data set or time series. Linear pattern of a lag plot ensures
non-random data and further suggests that an autoregressive model might be appropriate \cite{NIST}.
The lag plot of sample data give us an idea how well the data fit in ARIMA model.
The output of lag plot for the data of the group 1 is shown in the figure \ref{fig_lag_plot} and
the straight line behavior ensures that the data are going to fit an ARIMA model.

Now our next task is to find a suitable $ARIMA(p,d,q)$ model {\it i.e.}, the value of the parameters $p$, $d$, $q$
for the group 1 and group 2.
As the data of the group 1 are stationary then $d=0$, now we have to find $p$ and $q$ for respective the groups.
The algorithm is implemented on SPLUS software pack. It gives the best model to be an $ARIMA(2,0,30)$
with $p=2$ and $q=30$ often called $ARMA(2,30)$ for the data of group 1.
After doing similar statistical analysis we found that the data of the group 2 are also stationary {\it i.e.}, $d=0$ and
SPLUS software pack gives the best model $ARIMA(14,0,33)$ which is same as $ARMA(14,33)$ for the data of group 2 with $p=14$ and $q=33$.

\begin{figure}[h]
\centering
\includegraphics[width=3.5in]{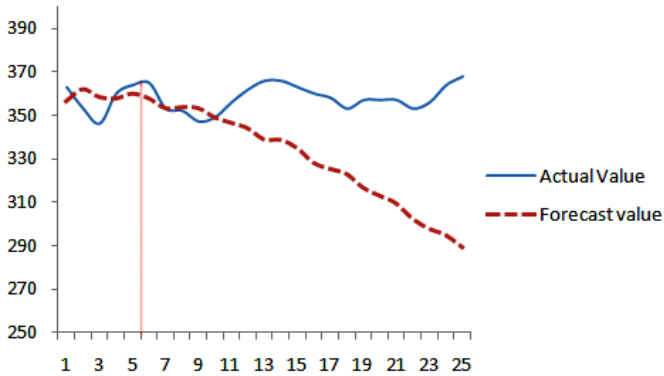}
\caption{Figure showing 25 step ARIMA forecast}
\label{fig_forecast}
\end{figure}

Above $ARIMA(2,0,30)$ model, we have used for forecasting and that forecast value can be used correcting data anomaly.
The figure \ref{fig_forecast} shows a 25 steps forecast of data for group 1. In the figure \ref{fig_forecast}, horizontal axis, {\it i.e.}, $x-$axis
represents forecast steps and vertical axis, {\it i.e.}, $y-$axis sample data.
\begin{figure}[h]
\centering
\includegraphics[width=3.5in]{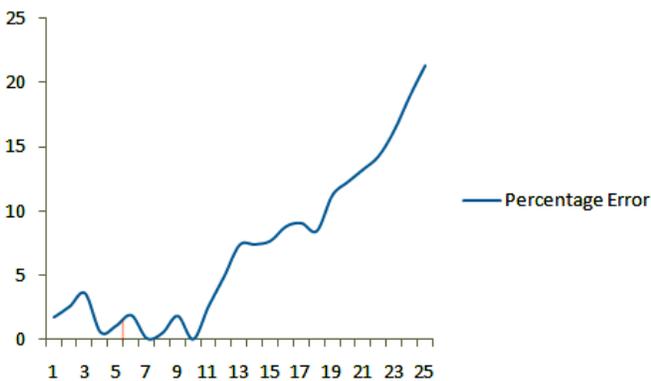}
\caption{Figure showing 25 step standard error for ARIMA forecast}
\label{fig_error}
\end{figure}
The corresponding error graph is shown in the figure \ref{fig_error} for verifying the notion that statistical
error accumulates during forecasting.
In this figure \ref{fig_error}, $x-$axis represents the corresponding forecast steps and $y-$axis represents the standard error.
It is clear from the figure \ref{fig_error} that the error increases with the number of forecast steps
after the first 10 step of forecast. The figure \ref{fig_error} suggests that we can use forecast values which are less than
the 10 step of forecast with insignificant error.

The table \ref{table:1} shows the calculation for first 5 steps of forecast values.
The actual values and the corresponding forecasted values are there in the table for comparison.
It is clear from the table that the maximum percentage error we make during fixing anomaly is 3.62\% only.
As stated above 5 steps are well enough to support our idea of using the model to 5 steps only.
\begin{table}[h]
\caption{Table showing first 5 steps of forecast values and corresponding errors}
\label{table:1}
%\begin{tabular}{|l|l|l|l|l|l|l|}
\begin{tabular}{lllllll}
  \hline
  \hline
  Forecast& Upper & Lower & Actual & Forecast & Forecast& Error \\
  Step & Bound & bound & Value & value & Std.err &\% \\
  \hline
    &     &	 &    &   &   &  \\
  1 & 369.32 &	343.90 &	363	& 356.61 &	6.48 &	1.75  \\
  &     &	 &    &   &   &  \\
  2 & 376.92&	347.29 &	353	& 362.10&	7.55&	2.58 \\
    &     &	 &    &   &   &  \\
  3 & 376.44&	340.64 &	346	& 358.54&	9.13&	3.62 \\
    &     &	 &    &   &   &  \\
  4 & 379.04&	336.67 &	360	& 357.85& 	10.80&	0.59 \\
    &     &	 &    &   &   &  \\
  5 & 384.79&	335.37 &	364	& 360.08&	12.60&	1.07 \\
  \hline
  \hline
\end{tabular}
\end{table}

\begin{figure}[h]
\centering
\includegraphics[width=3.5in]{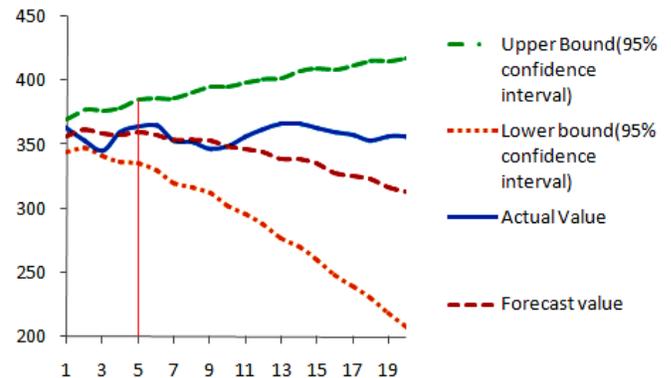}
\caption{Figure showing actual values, forecasted values and tighter upper and lower bounds for detecting anomaly}
\label{fig_bound}
\end{figure}

The same method, which is proposed in Algorithm \ref{alg:1} can be used to set the limits for detecting anomalous data.
We consider any data reported outside 95\% confidence interval as anomalous data and reject the data according
to the proposed Algorithm \ref{alg:2}. The figure \ref{fig_bound} shows the tighter upper and lower bound for
detecting anomaly, where $x-$axis represents forecast steps and $y-$axis sample data.
But, it is also clear from the same figure that the range of the bounds are increasing after the 5th
forecast step. In case of rejection due to data anomaly, a forecast value is generated by the Algorithm \ref{alg:1}
for replacement.

From the calculated data shown in the table \ref{table:1} it is clear that the maximum error we are tolerating
in rejecting correct value is 5\%. This 5\% is Type I error in our hypothesis testing stated in Algorithm \ref{alg:2}.
This 5\% values gives rise to 95\% confidence interval. Depending upon problem we can change the confidence interval
to find anomalies data.

As we change the confidence level for our experiment, the confidence interval get more narrow or broad.
If we want tighter bounds then we decrease the confidence level. As we keep on decreasing the confidence
level the size of interval gets smaller. But at the same time we will be loosing some good values as anomaly.
We also propose that if some node is showing repeated anomaly then it is necessary to check the node physically
for any permanent fault that may have been occurred.

\section{Conclusion}
The ARIMA model has been widely used for data modeling and prediction. It has ability to capture a wide
variety of realistic phenomena and is lightweight in terms of memory and computational cost. Its 
importance has not been till recognized by the research community of wireless sensor networks.
The data received by sink is often corrupt, missed, or dirty. In order to clean the data produced by WSN,
we developed a generalized framework using ARIMA model that identifies the degree of autocorrelation between
past data. To the best of our knowledge, this work is the first to utilize the ARIMA model for finding anomalies
within a stream of data for a single node and correct the anomalous data by appropriate forecast values.
The method protects the application from the abnormalities in the data by incorporating
aspects such as correlation and time of arrival of data. The novelty of this proposed method is that it provides
a mechanism that informs the system about the best suited smoothing process to be used for correcting anomaly.
We have validated our proposed method by using data from a real WSN application over the IRIS platform and demonstrated its
ability to detect and correct the anomalous data with quite a good accuracy. Future research work in this direction
includes the extension of our idea to detect node anomaly considering the spatial relationship among the
sensor nodes.

%The results show that data cleaning can be achieved
%without having any prior information about the input data stream and how correlation and time of
%arrival can be used for data cleaning.
\bibliographystyle{IEEEtran}
\bibliography{mybib}

\begin{thebibliography}{10}
\providecommand{\url}[1]{#1}
\csname url@samestyle\endcsname
\providecommand{\newblock}{\relax}
\providecommand{\bibinfo}[2]{#2}
\providecommand{\BIBentrySTDinterwordspacing}{\spaceskip=0pt\relax}
\providecommand{\BIBentryALTinterwordstretchfactor}{4}
\providecommand{\BIBentryALTinterwordspacing}{\spaceskip=\fontdimen2\font plus
\BIBentryALTinterwordstretchfactor\fontdimen3\font minus
  \fontdimen4\font\relax}
\providecommand{\BIBforeignlanguage}[2]{{%
\expandafter\ifx\csname l@#1\endcsname\relax
\typeout{** WARNING: IEEEtran.bst: No hyphenation pattern has been}%
\typeout{** loaded for the language `#1'. Using the pattern for}%
\typeout{** the default language instead.}%
\else
\language=\csname l@#1\endcsname
\fi
#2}}
\providecommand{\BIBdecl}{\relax}
\BIBdecl

\bibitem{Crossbow:2009}
``Crossbow technology inc.'' \emph{http://www.xbow.com/Products/products.htm},
  Accessed in February 2009.

\bibitem{Kaur:2010}
G.~Kaur, V.~Saxena, and J.~Gupta, ``Anomaly detection in network traffic and
  role of wavelets,'' in \emph{Proceedings of 2nd International Conference on
  Computer Engineering and Technology (ICCET 2010)}, vol.~7, april 2010, pp.
  46--51.

\bibitem{Ali:2010}
\BIBentryALTinterwordspacing
B.~Q. Ali, N.~Pissinou, and K.~Makki, ``Belief based data cleaning for wireless
  sensor networks,'' \emph{Wireless Communications and Mobile Computing}, 2010.
  [Online]. Available: \url{http://dx.doi.org/10.1002/wcm.970}
\BIBentrySTDinterwordspacing

\bibitem{Chong:2005}
C.~Liu, K.~Wu, and M.~Tsao, ``Energy efficient information collection with the
  {ARIMA} model in wireless sensor networks,'' in \emph{Proceeedings of Global
  Telecommunications Conference, IEEE GLOBECOM '05}, vol.~5, dec. 2005, pp. 5
  pp. --2474.

\bibitem{Jain:2004}
\BIBentryALTinterwordspacing
A.~Jain and E.~Y. Chang, ``Adaptive sampling for sensor networks,'' in
  \emph{Proceeedings of the 1st international workshop on Data management for
  sensor networks: in conjunction with VLDB 2004}, ser. DMSN '04.\hskip 1em
  plus 0.5em minus 0.4em\relax New York, NY, USA: ACM, 2004, pp. 10--16.
  [Online]. Available: \url{http://doi.acm.org/10.1145/1052199.1052202}
\BIBentrySTDinterwordspacing

\bibitem{Sutharshan:2007}
S.~Rajasegarar, J.~C. Bezdek, C.~Leckie, and M.~Palaniswami, ``Analysis of
  anomalies in {IBRL} data from a wireless sensor network deployment,'' in
  \emph{Proceedings of International Conference on Sensor Technologies and
  Applications (SENSORCOMM'07)}, Valencia, Spain, Oct, 2007, pp. 158--163.

\bibitem{Rajasegarar06distributedanomaly}
S.~Rajasegarar, C.~Leckie, and M.~Palaniswami, ``Distributed anomaly detection
  in wireless sensor networks,'' in \emph{Proceedings of Tenth IEEE
  International Conference on Communications Systems (IEEE ICCS 2006), 30
  October-1}, 2006.

\bibitem{Chitradevi:2011}
N.~Chitradevi, V.~Palanisamy, K.~Baskaran, and S.~Prabeela, ``Efficient
  distributed clustering-based anomaly detection algorithm for sensor stream in
  clustered wireless sensor networks,'' \emph{European Journal of Scientific
  Research}, vol.~54, no.~4, pp. 484--498, June 2011.

\bibitem{Box:1994}
G.~E.~P. Box and G.~M. Jenkins, \emph{Time Series Analysis: Forecasting and
  Control}, 3rd~ed.\hskip 1em plus 0.5em minus 0.4em\relax Upper Saddle River,
  NJ, USA: Prentice Hall PTR, 1994.

\bibitem{NIST}
\BIBentryALTinterwordspacing
``Handbook: Engineering statistics,'' 2011. [Online]. Available:
  \url{NIST/SEMATECH e-Handbook of Statistical Methods,
  http://www.itl.nist.gov/div898/handbook}
\BIBentrySTDinterwordspacing

\end{thebibliography}
\end{document}